\documentclass[aps,prd,floats,nofootinbib,preprintnumbers]{revtex4}
\usepackage{epsfig}
\usepackage{amsmath}
\usepackage{amssymb}
\usepackage{amsthm}

\begin{document}

\preprint{NUHEP-TH/07-06}

\title{GeV Seesaw, Accidentally Small Neutrino Masses, and Higgs Decays to Neutrinos}

\author{Andr\'e de Gouv\^ea}
\affiliation{Northwestern University, Department of Physics \& Astronomy, 2145 Sheridan Road, Evanston, IL~60208-3112, USA}

\begin{abstract}
If the Standard Model particle content is extended by gauge singlet fermions (right-handed neutrinos), active neutrinos generically acquire (Majorana) masses, in agreement with our current understanding of the lepton sector. If the right-handed neutrino masses are of the same order as the electroweak symmetry breaking scale, it is usually expected that these will not mediate any experimentally observable effects. Here, I explore the fact that this is not necessarily the case. If neutrino masses are ``accidentally small,'' active--sterile mixing angles can, according to current experimental bounds, be as large as several percent. If this is the case, I argue that the dominant decay mode of light ($M_H\lesssim 130$~GeV) Higgs bosons could be into an active and a sterile neutrino. The sterile neutrino decays promptly into a charge lepton and an on- or off-shell $W$-boson, so that the dominant Higgs boson decay mode is $H\to\ell+{\rm jets}~+$~missing transverse energy.

\end{abstract}

\maketitle

\setcounter{equation}{0} 
\setcounter{footnote}{0}
\section{Introduction}
\label{sec:intro}

Among the most important  particle physics results of the past twenty years is the discovery that neutrinos have, beyond reasonable doubt, nonzero masses and that leptons mix \cite{reviews}. Since the standard model (SM) predicts neutrino masses to exactly vanish, a ``new standard model'' ($\nu$SM) is required in order to properly describe the newly uncovered properties of the lepton sector. 

In spite of significant experimental and theoretical progress, we do not yet know what is the $\nu$SM. Several different options have been explored at length in the literature, and it is fair to say that a real breakthrough will only be achieved with the help of new experimental data. In this paper, I will concentrate on what is arguably the simplest version of the  $\nu$SM. It consists of the ``old'' SM with its matter sector augmented by gauge-singlet fermions $N_i$ ($i=1,2,\ldots$). Henceforth, I'll refer to the $N_i$ fields as right-handed neutrinos. The most general, renormalizable Lagrangian given the gauge symmetry and the field content is
\begin{equation}
{\cal L}_{\nu} \supset - y_{\alpha i} L_{\alpha}N_iH-\frac{1}{2}M_{ij} N_i N_j + H.c.,
\label{nuSM}
\end{equation} 
where $L_{\alpha}$ are lepton doublet fermion fields ($\alpha=e,\mu,\tau$), $H$ is the Higgs doublet scalar field, $y_{\alpha i}$ are neutrino Yukawa couplings, and $M_{ij}=M_{ji}$ ($i,j=1,2,\ldots$) are Majorana right-handed neutrino masses. 

After electroweak symmetry breaking, Eq.~(\ref{nuSM}) describes massive neutral Majorana fermions (neutrinos) whose masses are functions of $y_{\alpha i}$, $M_{ij}$ and $v=246$~GeV, the vacuum expectation value of the electrically neutral component of $H$. Current experimental data place only very modest bounds on $y_{\alpha i}$, $M_{ij}$: $10^{-12} \lesssim y_{\alpha i}\lesssim$~a few, while the magnitude of $M_{ij}$, can vary from $M_{ij}=0$, $\forall i,j$ (when all the  Majorana neutrinos ``fuse'' into Dirac neutrinos) to  $M_{ij}\sim 10^{16}$~GeV (when one runs into unitarity violation \cite{unitarity_violation}), excluding $M_{ij}\sim (0.001-0.1)$~eV, when strong active--sterile neutrino mixing, disfavored by current neutrino oscillation data, is expected. While the vast majority of studies of this Lagrangian in the literature \cite{Mohapatra:2004vr} concentrate on the limit where $M_{ij}\gg v$, it has recently been emphasized that all allowed values of $M_{ij}$ are technically natural, and should be considered as candidates for the $\nu$SM \cite{deGouvea:2005er}. In \cite{deGouvea:2005er,deGouvea:2006gz}, I, together with collaborators, have explored the phenomenological consequences of $M\lesssim1$~MeV. We uncovered non-trivial consequences for, {\it e.g.}, neutrino oscillation searches, precision measurements of tritium beta decay, searches for neutrinoless double-beta decay, and the detection of supernova neutrinos. It has also been pointed out in the literature that keV right-handed neutrinos could be the elusive dark matter \cite{Asaka:2005an,Asaka:2006ek}. 

For $M_{ij}\gtrsim 1$~GeV, one expects right-handed neutrinos to be outside the reach of nuclear, particle, and astrophysical probes. The reason for this is simple. Active\footnote{I refer to the neutrino mass eigenstates which are predominantly composed of $\nu_{e,\mu,\tau}$ as `active,' while the states that are predominantly composed of $N_{1,2,\ldots}$ are referred to as `sterile.'} neutrino masses $m_a$ are expected to be of order $y^2v^2/M$ where $y$ and $M$ are typical $y_{\alpha i}$ and $M_i$ values, respectively. On the other hand, typical active--sterile neutrino mixing angles, which govern the ``detectability'' of the mostly sterile states, are $\Theta\sim yv/M$. If the relations above are correct,  
\begin{equation}
\Theta\sim \sqrt{\frac{m_a}{M}} \lesssim 10^{-5}\left(\frac{\rm 1~GeV}{M}\right)^{\frac{1}{2}},
\label{general}
\end{equation} 
too small to lead to any observable effects. Here, I would like to emphasize that Eq.~(\ref{general}) does not necessarily apply, and that GeV seesaw neutrinos may indeed lead to interesting observable consequences. In Sec.~\ref{sec:formalism}, I describe the formalism summarized above in detail, and describe the circumstances under which Eq.~(\ref{general}) can be severely violated. I spell out conditions under which neutrino masses are accidentally small, and where $\Theta$ values can be as large as currently allowed by experiment. Current constraints are discussed in Sec.~\ref{sec:bounds}.  In Sec.~\ref{sec:future}, I discuss other consequences of GeV sterile neutrinos in the case of accidentally small neutrino masses, concentrating on the curious possibility that the  branching ratio for $H\to \nu_a N_i$ (where $H$ here is the propagating neutral Higgs boson) could be as large as that for $H\to b\bar{b}$. Under these circumstances, $N_i\to \ell W^{(*)}$, meaning that, a large fraction of the time, the Higgs boson decays into one charged lepton plus jets plus missing energy. In Sec.~\ref{sec:end}, I offer some comments and concluding remarks.

\setcounter{equation}{0}
\section{Yukawa Couplings, and Accidentally Small Neutrino Masses}
\label{sec:formalism}

Assuming the presence of $n$ right-handed neutrinos, after electroweak symmetry breaking, the $(3+n) \times (3+n)$ neutrino mass matrix is given by
\begin{equation}
m_{\nu}=\left(\begin{array}{cc} 0_{3\times 3} & yv \\ (yv)^t & M \end{array}\right),
\label{m_3+n}
\end{equation}
where here $y$ and $M$ are  $3\times n$ and $n\times n$ matrices, respectively, while $0_{3\times 3}$ stands for a three by three null matrix. Without loss of generality, I work on a weak basis where $M$ is diagonal, and all its eigenvalues are real and positive. The charged lepton mass matrix is also chosen diagonal, real, and positive. 

The $3+n$ Majorana neutrino masses are obtained by diagonalizing $m_{\nu}$ above. This will be done under the assumption that all elements of $(yv)M^{-1}$ are small. I refer to this assumption as the seesaw limit \cite{seesaw}. To leading order in  $(yv)M^{-1}$, the three lightest neutrino mass eigenvalues are given by the eigenvalues of 
\begin{equation}
m_a=yv M^{-1} (yv)^t,
\end{equation}
where $m_a$ is the mostly active neutrino mass matrix, while the $n$ heavy sterile neutrino masses coincide with the eigenvalues of $M$. The unitary matrix that diagonalizes $m_{\nu}$ ($U^t m_{\nu}U={\rm diag}(m_1,m_2,m_3,m_4\ldots,m_{(3+n)})$) is given by, at leading order in  $(yv)M^{-1}$,
\begin{equation}
U=\left(\begin{array}{cc} V & \Theta \\ -\Theta^{\dagger}V & 1_{n\times n} \end{array}\right),
\label{U_6x6}
\end{equation}
where $V$ is the unitary matrix that diagonalizes $m_a$
\begin{equation}
V^t m_a V = {\rm diag}(m_1,m_2,m_3), 
\end{equation}
$1_{n\times n}$ is the $n\times n$ matrix that (trivially) diagonalizes $M$, while
\begin{equation}
\Theta = (yv)^*M^{-1}.
\label{theta_y}
\end{equation}
Our current understanding of active neutrinos constrains $m_1,m_2,m_3$ and the elements of $V$, which is trivially related to the neutrino mixing matrix. Hence, we can express $\Theta$ in terms of active oscillation parameters, the sterile neutrino masses,  and a complex matrix $R$ that satisfies $RR^t=1$  by ``solving'' for $yv$ \cite{Casas:2001sr}:
\begin{eqnarray}
V^tyvM^{-1}(yv)^tV&=&{\rm diag}(m_1,m_2,m_3), \\
\left(V^tyvM^{-1/2}\right)\left(V^tyvM^{-1/2}\right)^t&=&\sqrt{{\rm diag}(m_1,m_2,m_3)}\sqrt{{\rm diag}(m_1,m_2,m_3)}, \\
V^tyvM^{-1/2} &=& \sqrt{{\rm diag}(m_1,m_2,m_3)} R, ~~~~~~{\rm if}~n\ge3~~~{\rm or}, \\
V^tyvM^{-1/2} &=& \sqrt{{\rm diag}(m_1,m_2,m_3)}R^t,~~~~~{\rm if}~n\le3.
\end{eqnarray}
In the case $n=3$, $R$ is an orthogonal matrix.\footnote{It is also important to keep in mind that in the $n<3$ case, some of $m_1, m_2,m_3$ vanish at this level.} For the case $n\le 3$, for example, one gets
\begin{eqnarray}
yv&=&V^*\sqrt{{\rm diag}(m_1,m_2,m_3)}R^tM^{1/2}, \\
\Theta&=&V\sqrt{{\rm diag}(m_1,m_2,m_3)}R^{\dagger}M^{-1/2}.
\end{eqnarray}

The point I want to emphasize here is that while neutrino masses are constrained to be very small, the elements of $\Theta$ can, in principle, be much larger than $\sqrt{m_i/M_j}$ ($i=1,2,3$, $j=1,\ldots,n$).\footnote{This point has been identified before in the literature \cite{Pilaftsis:1991ug,large_y}, mostly in the context of zero $m_a$ in the presence of non-zero $y$ and $M$. It has also been brought up recently in studies of weak scale resonant leptogenesis \cite{degenerate_leptogenesis,degenerate_leptogenesis_2}.} The reason for this is simple. While $R$ satisfies $RR^t=1$, its (complex) elements are in no way constrained to be small. Indeed, their magnitudes are, mathematically speaking, unbounded. 

It is illustrative to consider the case of one active neutrino of mass $m_3$ and two sterile ones, and further assume that $M_1=M_2=M$.  In this case, 
\begin{eqnarray}
\Theta&=&\sqrt{\frac{m_3}{M}}\left(\begin{array}{cc}\cos\zeta & \sin\zeta \end{array}\right), \\
yv&=&\sqrt{m_3M}\left(\begin{array}{cc} \cos\zeta^* & \sin\zeta^* \end{array}\right) \equiv \left(\begin{array}{cc}y_1 & y_2\end{array}\right),
\end{eqnarray}
where $\zeta\in \mathbb{C}$. If $\zeta$ has a large imaginary part, the magnitude of the elements of $\Theta$ is (exponentially) larger than $(m_3/M)^{1/2}$, while the neutrino Yukawa couplings are much larger than $\sqrt{m_3M}/v$, such that $m_3$ is much smaller than $y_1^2v^2/M$ or $y_2^2v^2/M$. The reason for this is a strong cancellation between the contribution of the two different Yukawa couplings to the active neutrino mass. For example, if $m_3=0.1$~eV, $M=100$~GeV, and $\zeta=14i$, the constraints above translate into $y_1\sim 0.244, y_2\sim -0.244i$, while $|y_1|-|y_2|\sim 3.38\times 10^{-13}$.

Many comments are in order.  Arbitrarily large values of $y$ and $\Theta$ are, of course, not allowed, for a couple of reasons. First, all results obtained above apply only to leading order in $(yv)M^{-1}$ and break down if $\Theta\sim 1$.\footnote{Of course, this possibility is experimentally ruled out.} Second, the self-consistency of the theory (the theory is required to be valid at least up to energies of order $M$) requires $y\lesssim$~a few. This means that the largest theoretically justified values of $\Theta$ which are still consistent with the seesaw approximation start to drop for $M\gtrsim 1$~TeV, {\it i.e.}, one is not allowed to violated $\Theta\lesssim v/M$. As an example, for $m_3=0.1$~eV $M=10^{12}$~GeV and $y_1\sim -iy_2<4$, the magnitude of the imaginary part of $\zeta$ is constrained to be less than 3, so that $\Theta$ values are constrained to be less than $10^{-10}$ --- only an order of magnitude larger than $\sqrt{m_3/M}$.

Another potential concern is the fact that the leading order results used above could be  completely inappropriate when the leading order active neutrino mass matrix is accidentally small, {\it i.e.}, when the elements of $m_a$ are much smaller than typical $(yv)^2/M$ values. Remarkably, the leading order results capture most of the exact results, as long as $(yv)M^{-1}$ is small. More to the point, it has been shown that, to all orders in $(yv)M^{-1}$, $m_a\propto yvM^{-1}(yv)^t$ \cite{seesaw_expansion}. 

Finally, one should worry about quantum corrections to all of the quantities in question ($y$ and $M$, for example). It seems clear that if $yvM^{-1}(yv)^t$ is accidentally small at the tree-level, it is expected to be much larger once one-loop corrections are included. While this is the case, nothing prevents $y$ and $M$ values such that the active neutrino masses are accidentally small at any order in perturbation theory. 

Before proceeding, I'll summarize the results of this section. I argued that, in the seesaw approximation, neutrino Yukawa couplings are {\sl not} constrained to be below $(4\times 10^{-8}) \sqrt{M/1~\rm GeV}$ so that active neutrino masses are below 0.1~eV. They can be orders of magnitude larger, as long as the active neutrino mass matrix is accidentally (very) small. In this case, as long as right-handed neutrino masses are much less than $10^9$~TeV, the active--sterile neutrino mixing angles are allowed to be much larger than naive expectations. This is so much so that, for right-handed neutrino masses below tens of TeV,  active--sterile neutrino mixing can  be considered as mostly independent from the active neutrino masses. In this spirit --- and ignoring ``naturalness'' concerns --- in the next two sections I'll discuss current constraints on the elements of $\Theta$, and will point out  interesting low-energy and collider consequences of the seesaw Lagrangian if the active neutrino mass matrix is accidentally small.

\setcounter{equation}{0}
\section{Constraints on Active--Sterile Mixing for GeV Right-Handed Neutrinos}
\label{sec:bounds}

In this section, I'll discuss the most stringent bounds on active--sterile mixing for weak scale right-handed neutrino masses: $M\in [5-200]$~GeV. The impact of significantly smaller right-handed masses was discussed in \cite{deGouvea:2006gz}, while some consequences of sub-GeV right-handed neutrinos have been recently studied in \cite{Gorbunov:2007ak}. Several of these constraints have been discussed in the literature in the context of ``generic'' sterile neutrinos \cite{sterile_const}, in the case of extra-dimensional neutrino mass models \cite{DeGouvea:2001mz,Ioannisian:1999cw}, or in the more general case of a non-unitary  neutrino mixing matrix \cite{Antusch:2006vw}. Some constraints on light right-handed neutrinos have also been discussed in the past
\cite{degenerate_leptogenesis_2,petcov,Chang:1994hz,Ilakovac:1994kj,Tommasini:1995ii,Ilakovac:1999md,Cvetic:2002jy,Loinaz:2003gc}. The results presented here update and combine different bounds discussed by different authors.

In more detail, I attempt to review the most stringent current bounds on several combinations of $U_{\alpha k}$, where $\alpha=e,\mu,\tau$, $k=4,\ldots,n+3$. These correspond to the  $\alpha i$ elements of $\Theta^*$, $i=1,\ldots,n$ (according to the definition of $U$ in Eq.~(\ref{U_6x6}), what is usually referred to as the active neutrino mixing matrix \cite{pdg} is $V^*$). 

The most stringent constraints on $U_{e k}$ come from failed searches for neutrinoless double beta decay. In the limit $M\gg 50$~MeV, the contribution of heavy, mostly sterile neutrinos can be approximately parameterized in terms of 
\begin{equation}
m_{ee}^{\rm heavy}=\sum_{i=1}^{n}\frac{U^2_{e (i+3)}}{M_i}Q^2,
\label{mee_heavy}
\end{equation}
where $Q^2\sim 50^2$~(MeV)$^2$. Naively, the fact that $m_{ee}\equiv m_{ee}^{\rm light}+m_{ee}^{\rm heavy}$ is constrained to be less than (roughly) 0.4~eV, translates into
\begin{equation}
U^2_{e(i+3)}\lesssim 2\times 10^{-7}\left(\frac{M_i}{\rm GeV}\right). 
\label{ue4_0nubb}
\end{equation}
The bound above is very uncertainty due to the fact that the relevant nuclear matrix elements associated to neutrinoless double beta decay are only poorly known \cite{Vogel:2006sq}. 

Care is required given that cancellations among the different terms in the sum may soften the constraint above significantly. As pointed out in \cite{deGouvea:2005er}, for example, if all right-handed masses are much less than the typical inverse size of the nuclei in question, the ``light'' and ``heavy'' contributions to $m_{ee}$ cancel, independent on whether the active neutrino mass matrix is accidentally small. 
Using Eq.~(\ref{theta_y}), Eq.~(\ref{mee_heavy}) can be written as
\begin{equation}
m_{ee}^{\rm heavy} =  \sum_{i=1}^{n}(yv)_{ei}^2 M^{-3}_i Q^2 . 
\label{Ue_bound}
\end{equation}
In the limit $M_i=M, \forall i$, $m_{ee}^{\rm heavy}=m^{\rm light}_{ee}Q^2/M^2$, where $m^{\rm light}_{ee}$ is the $ee$ element of  the active neutrino mass matrix, $m_a$ and the heavy contribution to $m_{ee}$ is negligible for values of $M$ above $\sim100$~MeV. Keeping such potential cancellations in mind, I'll stick to Eq.~(\ref{ue4_0nubb}) as the best constraint on the electron content of the mostly sterile neutrinos, unless otherwise noted.  

The active content of heavy, sterile neutrinos is also constrained by failed searches for neutrino oscillations at short baselines. In light of the strong constraints on $|U_{ek}|$ from neutrinoless double beta decay, the most relevant such constraint comes from searches for $\nu_{\mu}\leftrightarrow\nu_{\tau}$. The NOMAD experiment constrains $P_{\mu\tau}<1.7\times 10^{-4}$ at the 90\% confidence level in the limit of no baseline and neutrino energy dependency \cite{nomad_tau}. In the regime under consideration here, where it will turn out that the heavy neutrino states are (very) heavy and decay rather fast,\footnote{The relevant assumption here is that the muon neutrinos produced in the experiment are a linear combination of only the active mass eigenstates. This happens if the sterile states are too heavy and hence kinematically forbidden, or decay too fast so that they fail to reach the detector. If this is not the case, one must include the fact that the active and sterile states remain coherent and the constraint on active neutrino mixing is a factor of two more stringent.} $P_{\mu\tau}=|\sum_{i=1,2,3} U_{\mu i}U^*_{\tau i}|^2$, so that NOMAD constrains, at the 90\% confidence level,
\begin{equation}
\left|\sum_{k=4}^{3+n}U_{\mu k}U^*_{\tau k}\right|^2<1.7\times 10^{-4}.
\end{equation}
This constraint turns out to be less stringent than the ones discussed below.

Precision tests of the universality of charged current weak interactions provide other stringent constraints on active--sterile neutrino mixing. For example, the ratio of branching ratios for $\pi\to e\nu$ and $\pi\to \mu\nu$ is, assuming all sterile neutrino masses are larger than the pion mass, 
\begin{equation}
\frac{B(\pi\to e\nu)}{B(\pi\to \mu \nu)}\propto \frac{1-|\sum_{k=4}^{3+n}\left|U_{e k}\right|^2}{1-|\sum_{k=4}^{3+n}\left|U_{\mu k}\right|^2}\sim 1-\sum_{k=4}^{3+n}\left(\left|U_{e k}\right|^2-\left|U_{\mu k}\right|^2\right).
\end{equation}
Current measurements constrain, at the 90\% confidence level \cite{pdg,universality,pions}:
\begin{eqnarray}
\left|\sum_{k=4}^{3+n}\left(\left|U_{ek}\right|^2-\left|U_{\mu k}\right|^2\right)\right| & < 0.004,~~~(\pi~\rm decay) \label{umu4_pion} \\
\left|\sum_{k=4}^{3+n}\left(\left|U_{ek}\right|^2-\left|U_{\tau k}\right|^2\right)\right| & < 0.006,~~~(\tau~\rm decay) \\
\left| \sum_{k=4}^{3+n}\left(\left|U_{\mu k}\right|^2-\left|U_{\tau k}\right|^2\right)\right| & < 0.006,~~~(\tau~\rm decay) 
\end{eqnarray}
where I assume all right-handed neutrino masses to be larger than the tau mass. Similar constraints can also be obtained from kaon decays \cite{pdg,kaons}. Finally, there are two tantalizing hints for non-zero $U_{\alpha k}$, coming from the NuTeV anomaly and the invisible $Z$-width, indirectly measured at LEP \cite{pdg}. For a much more detailed discussion see, for example, \cite{nutev_nus}. The required $|U_{\alpha k}|^2$ are only marginally allowed by the bounds summarized above. 

Charged lepton flavor violating processes ($\mu\to e\gamma$, $\tau \to \ell \gamma$, $\mu\to eee$, $\mu-e$-conversion in nuclei, etc) are also sensitive to non-zero $U_{\alpha k}$. GIM suppression dictates that the rates for such processes (flavor changing neutral currents) grow with the neutrino mass, until $M\gtrsim M_W$, the $W$--boson mass. It is clear that, at leading order, the rate for charged lepton flavor violating muon processes is proportional to $|U_{e k}^*U_{\mu k}|^2$, while those for tau processes are proportional to  $|U_{\mu k}^*U_{\tau k}|^2$ or $|U_{e k}^*U_{\tau k}|^2$. 

\begin{figure}[htb]
\begin{center}
\includegraphics[scale=0.6]{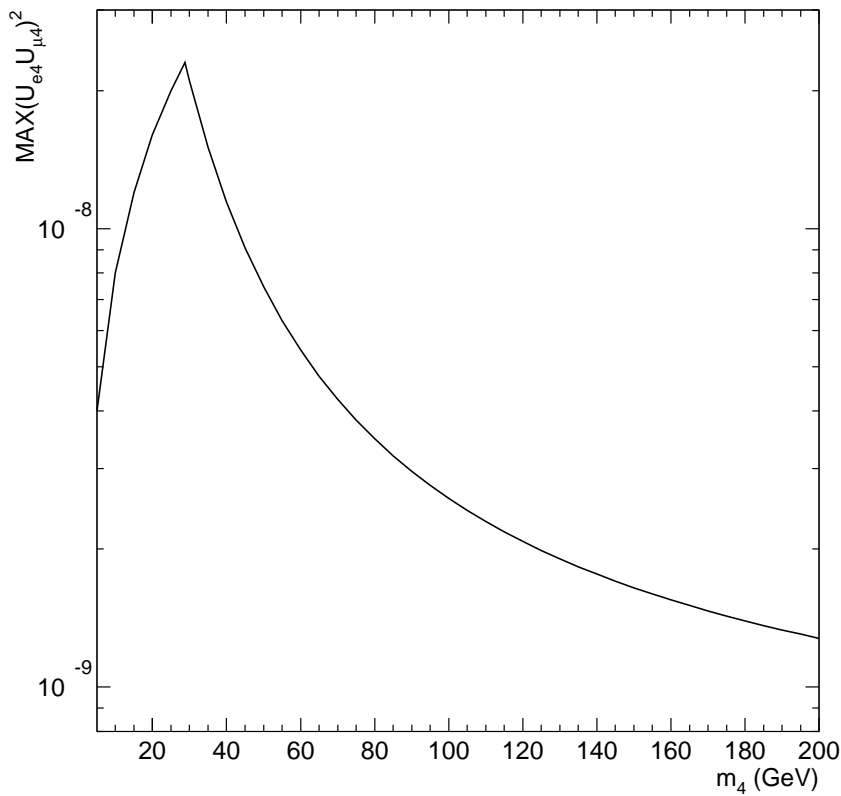}
\caption{Maximum phenomenologically allowed value for $|U_{e 4}^*U_{\mu 4}|^2$ as a function of $m_4$. At low masses, the most stringent constraints come from neutrinoless double-beta decay and tests of universality in pion decay, while at higher masses constraints from $\mu\to e$-conversion in $^{48}$Ti dominate.}\label{figure:bound}
\end{center}
\end{figure}
Assuming that only one sterile neutrino --- $\nu_4$ with mass $m_4$ --- is phenomenologically relevant  (the others are either heavier or more weakly coupled), the most stringent constraint on $|U_{e 4}^*U_{\mu 4}|^2$ comes from the non-observation of $\mu\to e$-conversion in nuclei \cite{Chang:1994hz}.  Fig.~\ref{figure:bound} depicts the maximum allowed value for $|U_{e 4}^*U_{\mu 4}|^2$ as a function of $m_4$. At low masses, the bound is dominated by the non-observation of neutrinoless double-beta decay , Eq.~(\ref{ue4_0nubb}),  combined with failed searches for universality violation in pion decay, Eq.~(\ref{umu4_pion}). At high masses, the dominant constraint comes from $B(\mu\to e$-conversion in $^{48}$Ti$)<4.3\times 10^{-12}$ at the 90\% confidence level \cite{pdg}.The result obtained here, which uses results obtained in \cite{DeGouvea:2001mz} (see appendix of \cite{DeGouvea:2001mz} for the detailed expressions), agrees qualitatively with the one computed in  \cite{Chang:1994hz}. 

For $m_4$ values in the electroweak  range, $|U_{e 4}^*U_{\mu 4}|^2$ cannot exceed a few times $10^{-8}$. Searches for charged lepton violating processes involving taus are not as stringent, and the most stringent constraints on $|U_{\delta4}^*U_{\tau4}|^2$ ($\delta=e,\mu$) come from tests of universality  in charged current interactions: $|U_{\mu4}^*U_{\tau4}|^2<2.4\times 10^{-5}$ at the 90\% confidence level, $|U_{e4}^*U_{\tau4}|^2\lesssim 1.2\times 10^{-9}(m_4/{\rm GeV})$. Given the strong constraints on $|U_{e 4}^*U_{\mu 4}|^2$, the limits above cannot be simultaneously saturated.

\begin{figure}[htb]
\begin{center}
\includegraphics[scale=0.6]{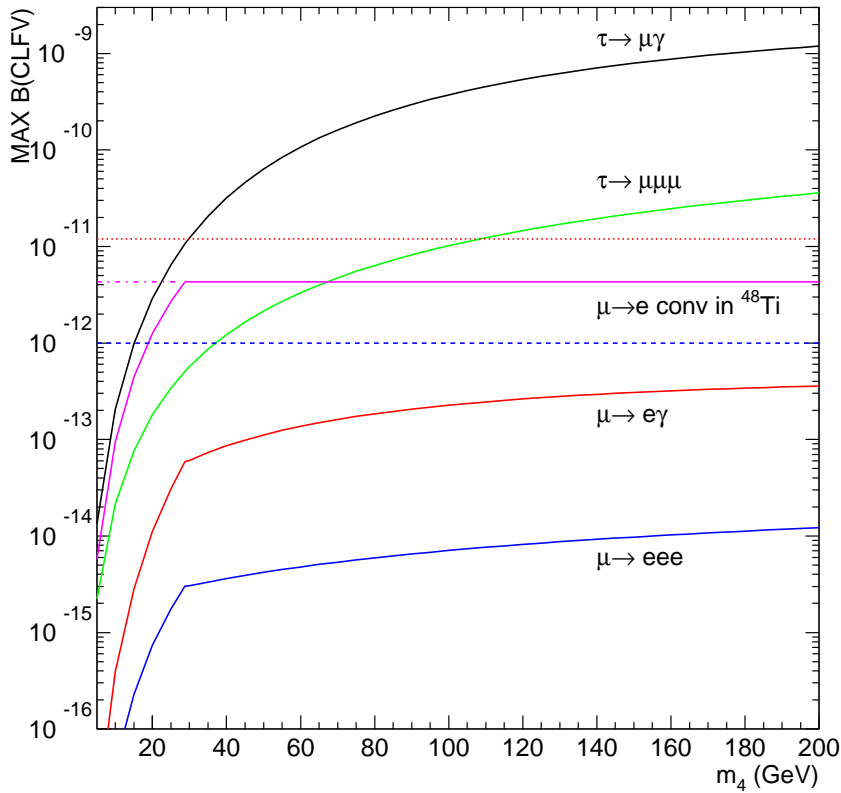}
\caption{Maximum allowed values for the branching ratios for selected charged lepton flavor violating processes (solid lines, from the top): $\tau\to\mu\gamma$ (black), $\tau\to\mu\mu\mu$ (green), $\mu\to e$-conversion in $^{48}$Ti (magenta), $\mu\to e\gamma$ (red), and $\mu\to eee$ (blue). The dotted, dashed-dotted, and dashed lines represent the current experimental bounds for $\mu\to e$-conversion in $^{48}$Ti (magenta), $\mu\to e\gamma$ (red), and $\mu\to eee$ (blue), respectively.}\label{figure:clfv}
\end{center}
\end{figure}

More sensitive searches for charged lepton flavor violation should explore much more significantly the $U_{\alpha 4}\times m_4$ parameter space, as long as $m_4\gtrsim 10$~GeV. Fig.~\ref{figure:clfv} depicts the maximum allowed values for the branching ratios for $\tau\to\mu\gamma$, $\tau\to\mu\mu\mu$, $\mu\to e$-conversion in $^{48}$Ti, $\mu\to e\gamma$, and $\mu\to eee$.

The MEG experiment at PSI \cite{meg} is expected to further constrain this scenario for $m_4\gtrsim 40$~GeV, while a next-generation $\mu\to e$-conversion experiment that reaches the sensitivity of the cancelled MECO experiment should further explore this scenario for all values of $m_4$ of interest. Significant bounds on all $|U_{\alpha 4}|$ can only be achieved with improved searches for charged lepton flavor violating processes involving taus, such as $\tau\to\mu\gamma$ and $\tau\to\mu\mu\mu$.  Currently, the B-factory experiments have ruled out $B(\tau\to\mu\gamma)>4.8\times 10^{-8}$ at the 90\% confidence level \cite{:2007vc} and $B(\tau\to\mu\mu\mu)>1.9\times 10^{-7}$ at the 90\% confidence level \cite{Yusa:2004gm}. It is expected that significantly better (at least one order of magnitude) sensitivity on $B(\tau\to\ell\ell\ell)$ will be achieved once larger fractions of the B-factory data samples are analyzed \cite{passaggio}. Yet more sensitivity (below $10^{-9}$?) will probably only be achieved with the advent of a Super-B factory \cite{Hewett:2004tv}. 

Finally, for ``large'' values of $|U_{\alpha 4}|^2$ and $m_4\lesssim 80$~GeV, very stringent constraints are imposed by searches for $Z\to \nu\nu_4$, followed by prompt $\nu_4\to \ell W^*$ or $\nu_4\to \nu Z^*$ \cite{Abreu:1996pa}. The branching ratio of  $Z\to \nu\nu_4$ is proportional to $\sum_{\alpha} |U_{\alpha 4}|^2$, and is constrained by DELPHI to be less than a few times $10^{-6}$ for $m_4\in[5,60]$~GeV. This translates into $\sum_{\alpha} |U_{\alpha 4}|^2$ less than a few times $10^{-5}$ in this mass window. Note that this constraint is not included in Figs.~\ref{figure:bound},\ref{figure:clfv}.

In summary, if only mixing with one sterile neutrino (say, $\nu_4$ with mass $m_4$) is relevant, the bounds above can be summarized as follows: $|U_{e4}|^2\lesssim 2\times 10^{-7}(m_4/~\rm GeV)$, $|U_{\mu4}|^2<4\times 10^{-3}$, $|U_{\tau4}|^2<6\times 10^{-3}$. The three bounds cannot be simultaneously saturated for $m_4\gtrsim 30$~GeV . For example, for $m_4=100$~GeV and $|U_{e4}|^2=2\times 10^{-5}$, $|U_{\mu4}|^2$ is constrained to be less than  $10^{-4}$. On the other hand, if both the $|U_{\tau 4}|^2$ and $|U_{\mu4}|^2$ bounds are saturated, $|U_{e4}|^2$ is constrained to be less than a few times $10^{-6}$ for all $m_4$ values. Furthermore, for $m_4\in[5,60]$~GeV, $\sum_{\alpha} |U_{\alpha 4}|^2\lesssim 3\times 10^{-5}$.

\setcounter{footnote}{0}
\setcounter{equation}{0}
\section{Collider Consequences of GeV right-handed neutrinos: $H\to \nu N$}
\label{sec:future}

Right-handed neutrinos with masses between a few and hundreds of GeV may play a role in collider experiments as long as their couplings to ``old'' SM particles are non-negligible. They mostly appear in processes involving on- and off-shell massive gauge bosons, with production cross-sections that are at least  $U_{\alpha k}^2$ times suppressed when compared to the production of active neutrinos. The most striking signature concerning the production of these sterile neutrinos is the potential violation of lepton number, as discussed in the literature \cite{lepton_at_collider,Han:2006ip}. For large active--sterile mixing angles, the Tevatron and the LHC should be sensitive to right-handed neutrino mediated lepton number violating processes, such as $pp\to \mu^+\mu^++X$ (where $X$ has no net lepton number). According to \cite{Han:2006ip}, for example, the LHC should ``see'' this lepton-number violating signal  for $m_4\sim 80$~GeV  as long $|U_{\mu 4}|^2\gtrsim 10^{-5}$ which, according to the results of the previous section, is currently allowed. 

\begin{figure}[htb]
\begin{center}
\includegraphics[scale=0.6]{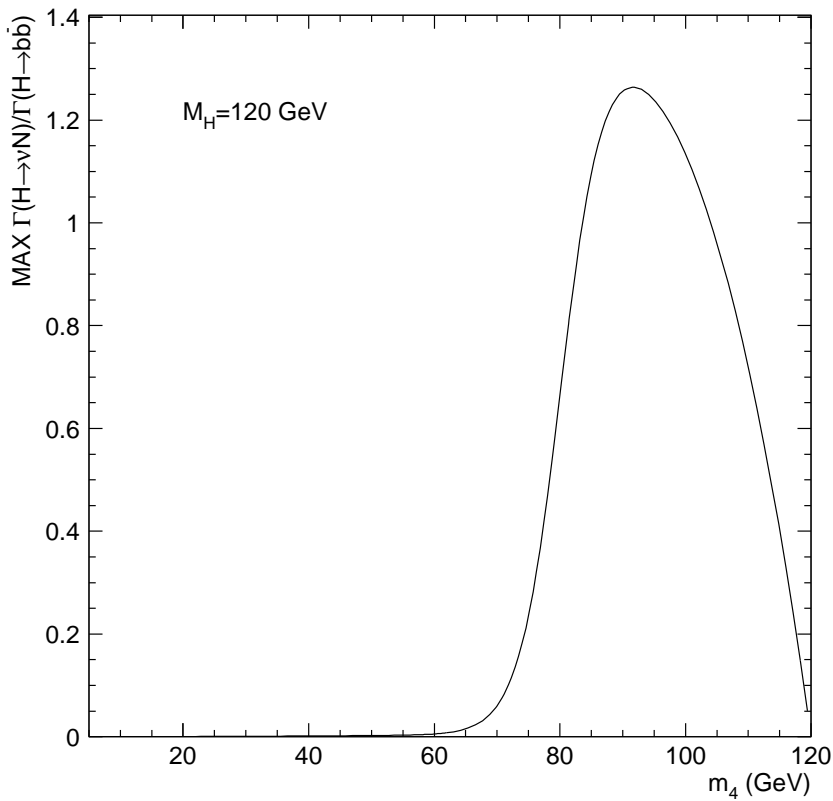}
\caption{Maximum allowed value of the Higgs boson decay width into neutrinos,  $\Gamma(H\to\nu_a \nu_4)$, in units of $\Gamma(H\to b \bar{b})$, as a function of $m_4$. The maximum value is computed assuming the upper bounds on active--sterile mixing discussed in Sec.~\ref{sec:bounds}.}\label{figure:higgs}
\end{center}
\end{figure}

Here I'd like to point out that, if the active--sterile mixing angles are close to their phenomenological upper bounds, the presence of right-handed neutrinos is expected to significantly modify the decay of the Higgs boson.  
This is easy to understand. Large active--sterile mixing for GeV right-handed neutrinos requires large neutrino Yukawa couplings, $y_{\alpha i}$. These control the partial width for $H\to\nu_{\alpha} N_i$, given by \cite{Pilaftsis:1991ug}  (with the approximation  $\sum_{i=1,2,3}|U_{\alpha i}|^2=1$),
\begin{eqnarray}
\Gamma(H\to \nu_{\alpha} N_i) &=& \frac{|y_{\alpha i}|^2}{8\pi}M_H\left(1-\frac{m^2_{N_i}}{M_H}\right), \\
&=& \frac{|U_{\alpha 4}|^2}{8\pi}M_H \frac{m_4^2}{v^2}\left(1-\frac{m^2_{4}}{M_H}\right),
\end{eqnarray}
for $N_i=\nu_4$. Fig.~\ref{figure:higgs} depicts the maximum allowed value of the Higgs boson decay width into neutrinos,  $\Gamma(H\to\nu N)=\sum_{\alpha}\Gamma(H\to \nu_{\alpha} \nu_4)$, in units of $\Gamma(H\to b \bar{b})$, assuming that the only relevant sterile state is $\nu_4$, as a function of $m_4$. Under these circumstances, Higgs decays --- which can also violate lepton number in a most invisible way --- constitute one of the leading production modes for right-handed neutrinos at colliders.

It is clear from the figure that, for a 120 GeV Higgs mass, the Higgs partial width into neutrinos can be of the same order of magnitude as the Higgs partial width into bottom quarks. On the other hand, the Higgs decay into neutrinos is far from invisible: at such large mixing angles, the $\nu_4$ state decays promptly as long as $m_4\gtrsim 4$~GeV. For $m_4\sim m_{\tau}$ (the tau mass), the lifetime $\tau_4$ of $\nu_4$ satisfies
\begin{equation}
c\tau_4=\left(\frac{m_{\tau}}{m_4}\right)^{5}\left(\frac{0.004}{|U_{\mu 4}|^2}\right)\times 21.8~\rm mm,
\end{equation} 
assuming $|U_{e4}|^2\ll |U_{\mu 4}|^2$. Taking this as a definitive upper bound to $c\tau_4$,\footnote{Of course, $c\tau_4$ scales not only like $(m_4)^{-5}$, but further decreases with $m_4$ as other decay modes open up, including $\nu_4\to \tau+$~jets and $\nu_4\to \tau+\ell\nu_{\ell}$, $\nu_4\to \mu+$heavy jets (charm and bottom), etc. For $m_4>M_W$, two-body decay modes like $\nu_4\to\mu+W$ open up, leading to much more prompt $\nu_4$ decays. Neutral current decays are also present but will not be discussed henceforth.} it is clear that for $m_4\sim5$~GeV and $|U_{\mu 4}|^2=0.004$, $\nu_4$ decays more promptly than $B$-mesons (by a factor of a few), while for  $m_4\sim30$~GeV and $|U_{\mu 4}|^2=0.004$, $\nu_4$ decays faster than neutral pions. For this reason the stringent LEP bound on failed searches for $Z\to\nu\nu_4$ with fast-decaying $\nu_4$ \cite{Abreu:1996pa} applies, and leads to the abrupt ``cut-off'' of the maximum allowed $\Gamma(H\to\nu\nu_4)$ observed in Fig.~\ref{figure:higgs}.

According to the discussion above, in the $\nu$SM, the invisible Higgs width is constrained to be small, {\it i.e.}, orders of magnitude below $H\to b\bar{b}$. On the other hand, the $H\to$~neutrinos partial width can be as large as $H\to b\bar{b}$, in which case this decay mode manifests itself at collider environments as $H\to E^{\rm miss}_t +\ell+$jets and $H\to E^{\rm miss}_t +\ell+\ell'$ (for $m_4<M_W$) or $H\to E_t^{\rm miss}+\ell +W$ (for $m_4>M_W$). Note that the $\nu$SM Higgs decays are {\sl not} the same as the more ``experimentally friendly'' $H\to NN$ decays discussed recently in the literature \cite{Graesser:2007yj,Graesser:2007pc}.\footnote{These also occur in the $\nu$SM as long as $M_i,M_j<0.5M_H$. Their rates are however, suppressed by an extra $|U_{\alpha k}|^2$ factor and, hence, negligible.} Most relevant is the fact that a purely sterile Higgs decay does not necessarily imply the presence of active neutrinos in the final state and hence missing transverse energy $E_t^{\rm miss}$. In \cite{Graesser:2007yj}, the contribution discussed here is neglected (which is safely the case if active neutrino masses are not accidentally small), and the Higgs decay to neutrinos is governed by an irrelevant operator ($\sim H^2N^2/\Lambda$). Nonetheless, some of the phenomenological considerations discussed in \cite{Graesser:2007pc} should also apply here. Large a large branching ratio for $H\to$~neutrinos can also be obtained in models with a fourth generation \cite{fourth}. In the scenario discussed, in \cite{fourth}, however, the fourth generation neutrino was stable.

If one assumes that the $H\to\nu N$ decay mode is not discernible, experimentally, from background, its main effect is to ``dilute''  the most promising Higgs decay channels at the Tevatron and the LHC. According to Fig.~\ref{figure:higgs}, the $\nu$SM branching ratio for $H\to b\bar{b}$ may be reduced by up to a factor of two with respect to the old SM one, rendering searches at the Tevatron somewhat more challenging. At the LHC, because the decay width of the light ($M_H\lesssim 130$~GeV) Higgs boson can be up to a factor of two larger, the branching ratio for $H\to\gamma\gamma$ should be similarly suppressed with respect to old SM expectations. If this turns out to be confirmed experimentally by the LHC experiments, it would be very important to investigate in more detail whether part of the Higgs decay width is being ``spent'' with the $\nu N$ final state, and tackle the experimental challenge of looking for $H\to \mu +W+E_t^{\rm miss}$ or, perhaps,   $H\to \mu^+ +e^-+E_t^{\rm miss}$. 

For smaller Higgs masses, currently disfavored by direct searches for the Higgs boson \cite{pdg}, the maximum allowed value for $B(H\to\nu N)/B(H\to b\bar{b})$ decreases with respect to the maximum value for the ratio of branching rations depicted in Fig.~\ref{figure:higgs}. For larger Higgs masses the behavior is just the opposite. For $M_H=200$~GeV, for example, the maximum allowed value for $B(H\to\nu N)/B(H\to b\bar{b})$ peaks at a little under four. For such heavy Higgs bosons, however, both branching ratios in question ($b\bar{b}$ and $\nu N$) are dwarfed by $H\to W^+W^-$ and $H\to ZZ$. Hence, the maximum impact of $H\to\nu N$  is expected for Higgs masses close to the current experimental lower bound (and in the region preferred by electroweak precision data). 

\setcounter{equation}{0}
\section{Comments and Conclusions}
\label{sec:end}

I pursued interesting phenomenological particle physics consequences of introducing gauge singlet ``right-handed neutrino'' fermions to the SM particle content. This scenario is arguably the simplest version of the $\nu$SM --- a Lagrangian capable of accommodating non-zero active neutrino masses --- and it has been argued in the past that the scale of right-handed neutrino masses is  embarrassingly unconstrained: almost all right-handed neutrino mass values between zero and $10^{16}$~GeV are phenomenologically allowed. Here, I concentrated on the region of parameter space where the values  of the right-handed neutrino Majorana masses coincide with the only other relevant $\nu$SM parameter, namely the Higgs mass-squared parameter that determines the weak scale.

Generically, expectations are such that quasi-sterile states are virtually invisible for right-handed neutrino masses above 1~GeV --- their couplings to the the visible SM sector are too feeble according to naive relations between neutrino masses and active--sterile mixing. I showed that it is possible to severely violate naive expectations by arguing that the entries of $yvM^{-1}(yv)^t$ could be ``accidentally small,'' while those of $yvM^{-1}$, which govern active--sterile mixing, could be close to several percent. If this turns out to be the case, active--sterile neutrino mixing ``decouples'' from the smallness of the neutrino masses, and is subject only to phenomenological bounds.

For right-handed neutrino masses above several GeV, the most stringent bounds on active--sterile mixing come from combinations of searches for neutrinoless double-beta decay, searches for universality violation in meson and lepton decays, and searches for charged lepton flavor violation. These bounds were summarized in Sec.~\ref{sec:bounds}. Future searches for rare muon processes --- especially $\mu\to e$-conversion in nuclei --- and tau processes should add significantly to our understanding of whether the seesaw scale is around the weak scale and whether neutrino masses are accidentally small.

For accidentally small neutrino masses and weak scale sterile neutrinos, I argued that sterile neutrinos may be ``copiously produced'' by Higgs decays, given the potentially large neutrino Yukawa couplings of weak scale right-handed neutrinos. It is remarkable that the branching ratio for $H\to \nu N$ can be larger than that for $H\to b\bar{b}$, the largest Higgs boson branching ratio for small Higgs masses in the old SM. If this is the case, the majority of Higgs bosons decay into $\nu\ell W^*$ or $\nu\ell W$ and will mostly manifest themselves as a charged lepton (muons or taus) plus jets (of all flavors) plus missing transverse energy. Such Higgs boson decay modes may prove a particularly daunting task for hadron collider  studies. 

Accidentally small neutrino masses and a weak scale seesaw energy scale also allow for large lepton number violating effects at the Tevatron and LHC, as discussed in some detail in the literature. Here, I would like to point out that even if one attempts to go outside of the $\nu$SM in order to include new, strongly-mixed, weak scale quasi-sterile neutrinos, the ``accidentally small'' discussion in Sec.~\ref{sec:formalism} still applies. This is easy to see. Any sterile neutrino that directly mixes with the active neutrinos and has a much larger Majorana mass will induce, via the seesaw mechanism, a contribution to the neutrino mass matrix given by, schematically,  $yvM^{-1}(yv)^t$, while the active--sterile mixing angle is given by $yvM^{-1}$. Some form of cancellation --- usually very severe --- is {\sl required} in order to allow for large active--sterile mixing in light of our current understanding of active neutrino masses. The cancellation need not be between different contributions to $yvM^{-1}(yv)^t$. For example, the sterile neutrino contribution to the active neutrino mass may be cancelled by a ``Type-II'' seesaw contribution (for a recent related discussion, see \cite{Akhmedov:2006de}). 

It is interesting to ask whether the ``accidental'' aspect of the accidental neutrino mass scenario described in Sec.~\ref{sec:formalism} can be explained by a hidden horizontal symmetry. One possibility would be a symmetry that guaranteers, at leading order, that $m_a$ vanishes for non-zero $M$ and $y$ (see discussions in \cite{Pilaftsis:1991ug,large_y}).  Breaking effects would have to be kept small in order to naturally explain the large Yukawa couplings that saturate the constraints described here.  A detailed discussion of this possibility within a specific flavor-symmetry scenario was very recently presented in \cite{Kersten:2007vk}.

{\bf Note Added}: During the completion of this work, \cite{Kersten:2007vk} became available in the preprint archives. The authors of \cite{Kersten:2007vk} also address the requirements for ``accidentally small'' neutrino masses (concentrating more on zero leading order neutrino masses) and some of their discussions are similar to the one in Sec.~\ref{sec:formalism}. While  \cite{Kersten:2007vk} concentrates on obtaining a natural understanding for the accidentally small neutrino masses and understanding its consequences, here I concentrate more on phenomenological bounds on accidentally small neutrino masses, and experimental consequences of a scenario where these bounds are saturated. Finally, small preliminary subsets of the results discussed here were presented earlier at the Pheno 2007 Symposium \cite{pheno} and FPCP 2007 Conference \cite{fpcp}.

\section*{Acknowledgments}

I would like to thank Boris Kayser and Neal Weiner for inspiring discussions, and Radovan Derm\'{\i}\v{s}ek for words of encouragement. I am also indebted to Joern Kersten for pointing out the very relevant bounds from LEP searches for neutral heavy leptons, which were not included in an earlier version of this manuscript. I am happy to acknowledge the hospitality and very productive environment of the Aspen Center for Physics, where the final stages of this work were completed. This work is sponsored in part by the US Department of Energy Contract DE-FG02-91ER40684.

 \end{document}